\documentclass[twocolumn,showpacs,preprintnumbers,amsmath,amssymb]{revtex4}

\usepackage{wasysym}
\usepackage{epsfig}
\usepackage{graphicx}
\usepackage{dcolumn}
\usepackage{bm}
\usepackage{verbatim}

\newcommand{\om}{\omega}
\newcommand{\Om}{\Omega}

\newcommand{\be}{\begin{equation}}
\newcommand{\ee}{\end{equation}}
\newcommand{\bea}{\begin{eqnarray}}
\newcommand{\eea}{\end{eqnarray}}

\begin{document}

\preprint{APS/123-QED}

\title{High-resolution atom interferometers with suppressed diffraction phases
}

\author{Brian Estey}
\author{Chenghui Yu}
\author{Holger M\"uller}
\altaffiliation{Lawrence Berkeley National Laboratory, One Cyclotron Road, Berkeley, California 94720, USA}

\email{hm@berkeley.edu}

\affiliation{Department of Physics, 366 Le Conte Hall MS 7300, University of California, Berkeley, California 94720, USA}
\author{Pei-Chen Kuan}
\author{Shau-Yu Lan}
\affiliation{Division of Physics and Applied Physics, School of Physical and Mathematical Sciences, Nanyang Technological University, Singapore, 637371 Singapore}
\date{\today}

\begin{abstract}
We experimentally and theoretically study the diffraction phase of large-momentum transfer beam splitters in atom interferometers based on Bragg diffraction. We null the diffraction phase and increase the sensitivity of the interferometer by combining Bragg diffraction with Bloch oscillations. We demonstrate agreement between experiment and theory, and a 1500-fold reduction of the diffraction phase, limited by measurement noise. In addition to reduced systematic effects, our interferometer has high contrast with up to 4.4 million radians of phase difference, and a resolution in the fine structure constant of $\delta \alpha/\alpha=0.25\,$ppb in 25 hours of integration time.
\end{abstract}
\maketitle

Atom interferometers are a direct analogy to optical interferometers, where beam splitters and mirrors send a wave along two different trajectories. When the waves are recombined, they can interfere constructively or destructively, depending upon the phase difference $\Delta \phi$ accumulated between
the paths. In light-pulse atom interferometers, atomic matter waves are coherently split and reflected using atom-photon interactions, which impart photon momenta $\hbar k$ to the atoms.

\begin{figure}
\centering
\epsfig{file=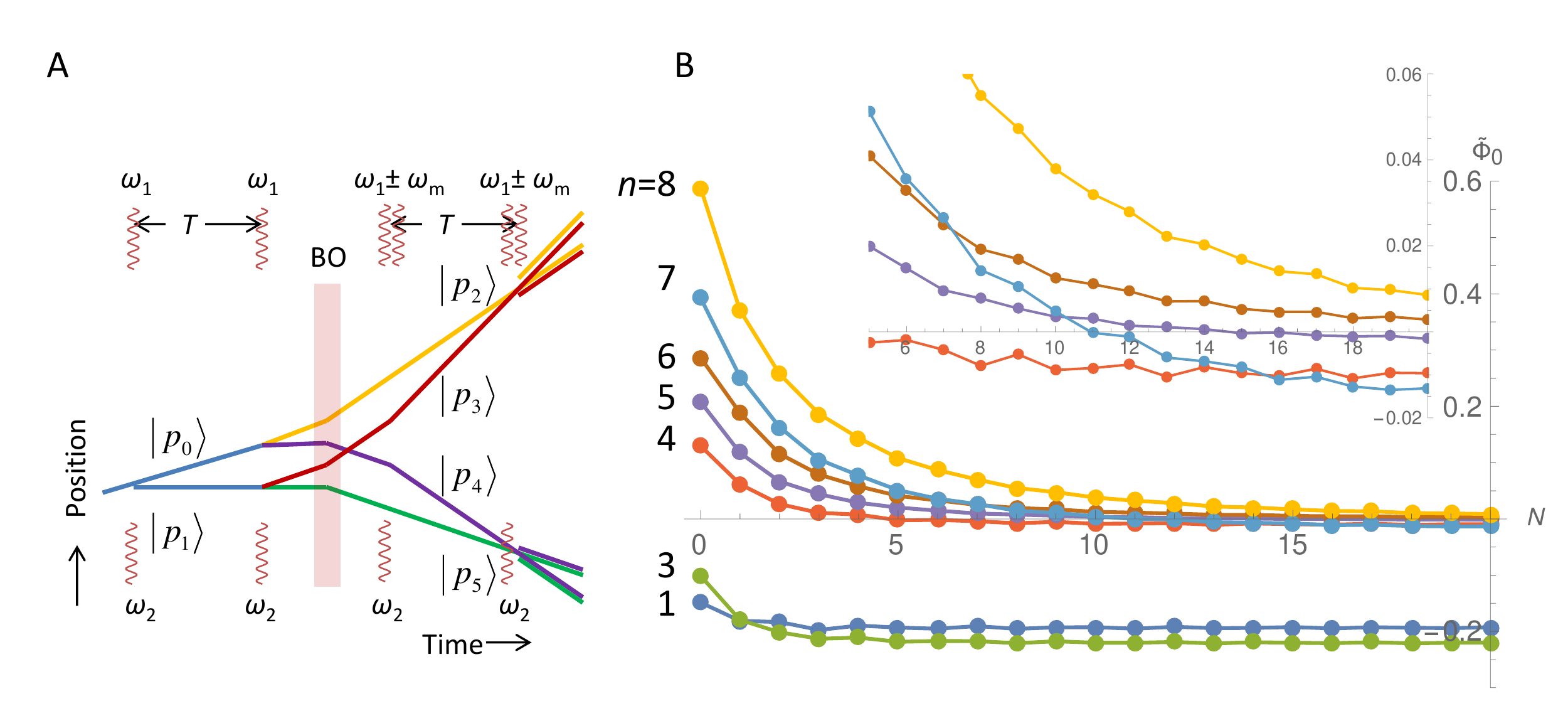,width=0.5\textwidth}
\caption{\label{RBI} A: Trajectories of our simultaneous conjugate interferometers with Bloch oscillations. Gravity has been neglected. The wiggly lines indicate laser pulses driving Bragg diffraction, the shaded area marked "BO" are the Bloch oscillations. B: Strong suppression of the beam splitter phase shift in radians as function of the number $N$ of Bloch oscillations for Bragg diffraction orders of $n=1-8$ ($n=2$ lies outside the scale at $\tilde\Phi_0\sim -0.6$). The inset shows an enlarged part for large Bloch oscillation numbers $N$.}
\end{figure}

In a Ramsey Bord\'e interferometer, for example, the atom (of mass $m$) moves away and back along one path while remaining in constant inertial motion along the other path. The phase difference $\Delta \phi=8\om_r T$ ($T$ is the pulse separation time) is proportional to the kinetic energy, and thus to the recoil frequency $\om_r=\hbar k^2/(2m)$. This enables
state-of-the-art measurements of the fine structure constant $\alpha$ \cite{Bouchendira2011} and will help realize the expected new definition of the kilogram in terms of the Planck constant \cite{CCC,Bouchendira2013}. Using multiphoton Bragg diffraction \cite{Losses,BraggPRL} and simultaneous operation of conjugate interferometers \cite{SCI}, the phase difference has been increased to $\Phi=16n^2\om_r T$ (where the factor of 16 arises from taking the phase difference of the two interferometers), and Earth's gravity and vibrations have been canceled. 
Unfortunately, however, Bragg diffraction causes a diffraction phase \cite{Cronin,Buechner,CCC,Jamison}, which has been the largest systematic effect in high-sensitivity atom interferometers using this technique \cite{CCC}. 
Here, we study the diffraction phase in detail and show that it can be suppressed and even nulled by introducing Bloch oscillations as shown in Fig. \ref{RBI} A, B. Bloch oscillations also increase the measured phase shift to
\be\label{omm}
\Phi=16n(n+N)\om_r T,
\ee
where $n>1$. 
We decrease the influence of diffraction phases by an amount that is considerably larger than the increase in sensitivity and are in fact able to null them by feedback to the laser pulse intensity. With this increase in signal and suppression of diffraction phase systematics, we expect to see improvements in many applications of atom interferometry, such as measuring gravity and inertial effects \cite{
Geiger,LVGrav,Altin,Sugarbaker}, measuring Newton's gravitational constant $G$ \cite{Fixler,TinoG}, testing the equivalence principle \cite{
redshift,redshiftPRL,nuclear,EEP,Poli,GravAB,Dickerson}, CPT and Lorentz symmetry \cite{Stadnik} and perhaps even antimatter physics \cite{antiHint} and detecting gravitational waves \cite{GravWav}.

Diffraction phases occur between the waves reflected and transmitted by a beam splitter and cause an unwanted (usually) shift of the interference pattern in an interferometer. In light-pulse atom interferometers with Raman beam splitters, a pair of laser beams drive two-photon transitions between two hyperfine states. Here, diffraction phases are caused by differential ac Stark effects between these states and can be nulled by choosing a certain intensity ratio of the laser beams \cite{Weiss}. While these can be several radians large, Raman systems are very nearly perfect two-level systems, i.e., the atom can remain in the original state or be transferred, but not lost to a third state. The last half of the atom interferometer is therefore a time-reversed mirror image of the first half, and diffraction phases cancel unless the symmetry is broken by technical imperfections. (Atoms undergoing incoherent single-photon events don't interfere and thus don't cause phase offsets.) In Bragg diffraction, differential Stark shifts are absent because the atoms do not change their internal quantum state. Coherent coupling of the atoms to unwanted momentum states, however, creates diffraction phases and causes atom loss, breaking the time-reversal symmetry of the interferometer. Thus, these diffraction phases are not easily canceled. They depend on the duration, intensity, and shape of the laser pulses, on the Bragg diffraction order, as well as on the detuning from two-photon resonance and thus on the atom's velocity (but are nonzero even for zero velocity). In simultaneous conjugate interferometers, they also depend on the detuning $2\om_m$ between the frequencies used for the final two pulses. 

The main building blocks of our interferometer are Bragg diffraction and Bloch oscillations. During Bloch oscillations, a matter wave is loaded into an optical lattice created by counterpropagating laser beams \cite{Peik,CladeBloch}. When the optical lattice is accelerated by ramping the frequency difference of the lasers, the atom's velocity expectation value will follow the acceleration of the lattice, in addition to oscillating around its mean value with a period of $\tau_B=8\om_r/\dot \om$. If the lattice is turned off when a momentum of $2N\hbar k$ has been transferred to the atom, where $N=1,2,3,\ldots$, the atom is found in a pure momentum state.

In Bragg diffraction \cite{Losses,BraggPRL}, an atom absorbs $n$ photons of momentum $\hbar k_1$ and is stimulated to emit $n$ photons of momentum $\hbar k_2$ into the opposite direction, receiving a total impulse of $2n\hbar k$, where $ k= (k_1+k_2)/2$. The atom's internal state remains unchanged. 
In our experiment (Fig. \ref{RBI} A), 
the first beam splitter transfers an impulse of $2n\hbar k$ with 50\% probability, driven by two laser frequencies $\omega_{1,2}$. After an interval $T$, a second beam splitter forms two pairs of parallel-moving paths. The upward-moving pair is accelerated further by $2N\hbar k$ using Bloch oscillations; the downward-moving pair is accelerated down simultaneously, by the same amount. This requires two optical lattices accelerated in opposite directions. 
After the acceleration, two more beam splitters close the two interferometers.

Since no single Bragg diffraction or optical lattice can be simultaneously resonant for the upper and lower interferometer, we use three laser beams. To this end, the beam at frequency $\om_1$ is replaced by two, equally strong, beams at $\om_1^\pm$, where
\be
\om_m\equiv \frac{\om_1^+-\om_1^-}{2}, \quad \om_1\equiv \frac{\om_1^++\om_1^-}{2}, \quad \om_{12}\equiv \om_1-\om_2.
\ee
Accounting for the the two frequencies $\om_1^\pm$ adds a phase term $-2\om_m n T$ to Eq. (\ref{omm}) via laser-atom interaction \cite{CCC}. A measurement of the recoil frequency can proceed by changing the frequency $\om_m$ until $\Phi=0$; This leads to
\be\label{ommeas}
\frac{\om_m}{8(n+N)}=\frac{\tilde\Phi_0}{16n(n+N)T}+\om_r \equiv \frac{\Phi_0}{T}+\om_r,
\ee
where we call $\tilde \Phi_0$ the diffraction phase and $\Phi_0=\tilde \Phi_0/[16n(n+N)]$ the reduced diffraction phase.

To study the Bragg beam splitters in detail, we consider an atom in the light field of the three frequencies $\om_1^\pm$ and $\om_2$, using the rotating wave approximation. Since all frequencies are far-detuned from any atomic transition, we may adiabatically eliminate the excited state. We expand the ground-state wave function $g(z,t)$ in momentum states $g(z,t)=\sum_m \exp(-4i m^2\om_r t+2imkz)g_m(t)$. The Schr\"odinger equation reads
\bea\label{sgl}
i\dot g_m=\frac{\Om_R}{2}2\cos(\om_m t)\left(e^{-i\om_{12}t+4i(2m-1)\om_r t}g_{m-1}\right. \nonumber \\ \left. +e^{i\om_{12}t-4i(2m+1)\om_rt}g_{m+1}\right).
\eea
If we set $2\cos(\om_m t)$ to one, we recover the equations for the first two Bragg pulses, which are driven by $\om_1, \om_2$ only. 
For the Rabi frequency, we assume a gaussian time-dependence, $\Om_R=\hat \Om_R e^{-t^2/2\tau^2}$. Numerically solving these equations yields the matrix elements $\langle m|a, b|n\rangle $ which give the amplitudes for the Bragg pulse to transfer an atom from a momentum state $|n\rangle$ moving at $2n \hbar k$ into a momentum state $|m\rangle$ when driven with laser frequencies $\om_{12}=8a \om_r$ and $\om_m=8b\om_r$. We also denote $\langle m|a|n\rangle$ as the corresponding amplitude when there are only two Bragg frequencies. The matrix elements have the symmetries $\langle m|a,b|n\rangle=\langle n|a,b|m\rangle =\langle-m|-a,b|-n\rangle  =\langle n+c|a+c,b|m+c\rangle$. For states $n,m$ that satisfy Bragg resonance, we have furthermore ${\rm arg}(\langle n|n|n\rangle)={\rm arg}(\langle0|n|0\rangle)$.

Labeling the momentum states as in Fig. \ref{RBI}, the probabilities $|\psi_{1-4}|^2$ for an atom to arrive in the four outputs (counting from top to bottom in Fig. \ref{RBI} A) can be computed from the evolved states, which read, for example,
\bea
\psi_1=|2n+N\rangle \nonumber \\ \times ( \langle p_3|n,n+N|p_2\rangle \langle p_2|n,n+N|p_2\rangle \langle p_0|n|p_0\rangle^2 \nonumber \\ +
ie^{i\phi_1} \langle p_3|n,n+N|p_3\rangle \langle p_3|n,n+N|p_2\rangle\,\langle p_0|n|p_1\rangle^2 ),\quad
\eea
where $\phi_1$ is any phase difference between the two paths not arising from diffraction phases, and we assume the atom entered in a state $|p_0 \rangle$.

In order to obtain the diffraction phase shift, we write $|\psi_1|=|c_1+e^{i\phi_1+i\tilde\phi^0_1}c_2|$, where $c_{1,2}$ and $\tilde\phi^0_1$ are real constants. The latter is the diffraction phase measured by detecting the population $|\psi_1|^2$ at this interferometer output. Similar diffraction phases $\tilde\phi^0_{2-4}$ are obtained from the other three outputs. Experimentally, we use normalized detection and measure $(|\psi_{1}|^2-|\psi_{2}|^2)/(|\psi_{1}|^2+|\psi_{2}|^2)$. The diffraction phase thus measured is $(\tilde\phi^0_1-\tilde\phi^0_2)/2$ and the phase of the interferometer pair is $\tilde \Phi_0=(\tilde\phi^0_1-\tilde\phi^0_2+\tilde\phi^0_3-\tilde\phi^0_4)/2$.

While the physics of the Schr\"odinger equation Eq. (\ref{sgl}) is complex, a few observations can be made. Phase shifts caused by the first beam splitter (Fig. \ref{RBI}) cancel exactly. Also, for $n=4, N=25$, phase shifts are most sensitive to the parameters (e.g., intensity) of the second and third beam splitter, while the influence of changes to the last beam splitter is suppressed by almost three orders of magnitude. It is currently unknown whether this holds generally for any values of $n, N$. Finally, we note that the diffraction phase can be exactly canceled in Mach-Zehnder interferometers: 
No diffraction phase will be measured when detecting only the output state moving with the momentum of the incoming atom. 

Figure \ref{RBI} B shows how the calculated diffraction phase is strongly decreased by introducing Bloch oscillations. This suppression arises when introducing a large number of Bloch oscillations, which leads to a large Doppler effect between the interferometers at the time of the third and fourth beam splitter. The light-atom interaction in each interferometer becomes less influenced by the laser frequency intended to address the  other interferometer. This suppression of the absolute diffraction phase is augmented by a signal increase by a factor of $16n(n+N)$ which further reduces the relative influence.

\begin{figure}
\centering
\epsfig{file=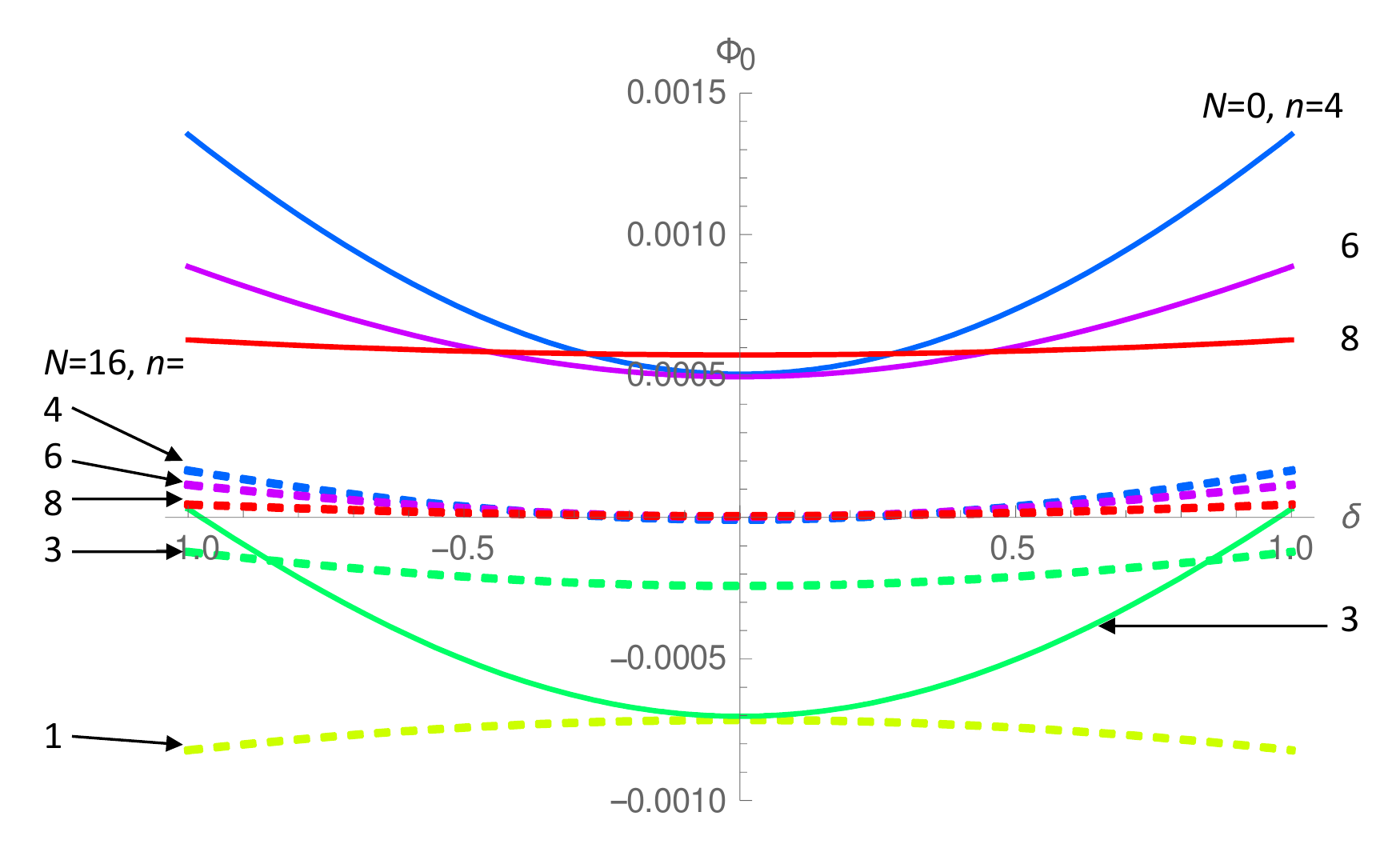,width=0.45\textwidth}
\caption{\label{SCInew} Reduced diffraction phase shift in radians calculated for a pair of interferometers as function of the detuning $\delta/\om_r$ from the Bragg resonance. The solid lines indicate the phase without Bloch oscillations, $N=0$, dashed lines with $N=16$. The nominal pulse duration is $\tau=0.2/\om_r$, with the pulse peak intensity adjusted so as to provide 50\% diffraction efficiency. The Gaussian pulses are truncated at 1/40 of their peak amplitude. 
}
\end{figure}

Figure \ref{SCInew} shows the reduced diffraction phase as function of the two-photon detuning $\delta$ of $\om_{12}$ from Bragg resonance, and therefore on atom velocity. 
Thermal atoms have a velocity distribution which leads to a distribution of the two-photon detuning through the Doppler effect. To model this, we assume an initially flat distribution from which atoms are selected by two Doppler-sensitive Raman transitions driven by square-shaped $\pi$-pulses of duration $t_s$. 
Table \ref{BSphase} lists the theoretical diffraction phases $\tilde \Phi_0$. For $n=5$, e.g., we obtain a $17$\,mrad phase shift at $N=16$ after integrating over the velocity distribution of a $400\,\mu$s velocity selection pulse. This is an improvement of about 13 times relative to the case of no Bloch oscillations, even before the reduced diffraction phase and thus the increased signal size are considered. A further reduction and reduced sensitivity to the two-photon detuning can be achieved by increasing the pulse duration \cite{supplement}.

\begin{table}
\centering
\caption{\label{BSphase} Diffraction phases fitted as $\tilde\Phi_0=\varphi_0+\varphi_2 (\delta-\delta_0)^2$. The table lists $\varphi_{0,2}$ as function in mrad for $N=0, 16$ and $n=1,\ldots, 10$. $|\delta_0|<10^{-8}$ for all cases. The phase $\tilde \Phi_0$ has to be divided by $16 n(n+N)$ to obtain the reduced diffraction phase $\Phi_0$. Throughout, $\tau=0.2$.}
\begin{tabular}{cc|cccccccccc} \hline\hline
$N$  & & $n=1$ & 2 & 3 & 4 & 5 & 6 & 7 & 8 & 9 & 10 \\ \hline
0 & $\varphi_0$ & -149 & -720 & -100 & 132 & 210 & 289 & 400 & 588 & 754 & 840 \\
  & $\varphi_2$ & -19.8 & 13.8 & 106 & 218 & 193 & 225 & 237 & 54.5 & 40 & 165 \\
16 & $\varphi_0$ & -192 & -826 & -220 & -7.7 & 2.2 & 7.5 & -3.9 & 16.3 & 24.8 & -3.8 \\
   & $ \varphi_2$ & -29.1 &  9.0 & 111 & 223 & 209 & 239 & 269 & 127 & 125 & 249 \\ \hline\hline
\end{tabular}
\end{table}

Our experiment is performed in a $\sim 1$-m high atomic fountain of cesium atoms at $0.4\,\mu$K in the $m_F=0$ state, obtained by a moving-molasses launch and three-dimensional Raman sideband cooling in a moving optical lattice. 
Two vertical velocity cuts with a pulse duration $t_s$ of 400$\,\mu$s are performed. Signals due to gravity and vibrations are cancelled by measuring the phase difference of the two interferometers using ellipse fitting \cite{SCI}. The laser system driving the interferometer is similar to the one described in \cite{BraggPRL,SCI,BBB}.

\begin{figure*}
\centering
\epsfig{file=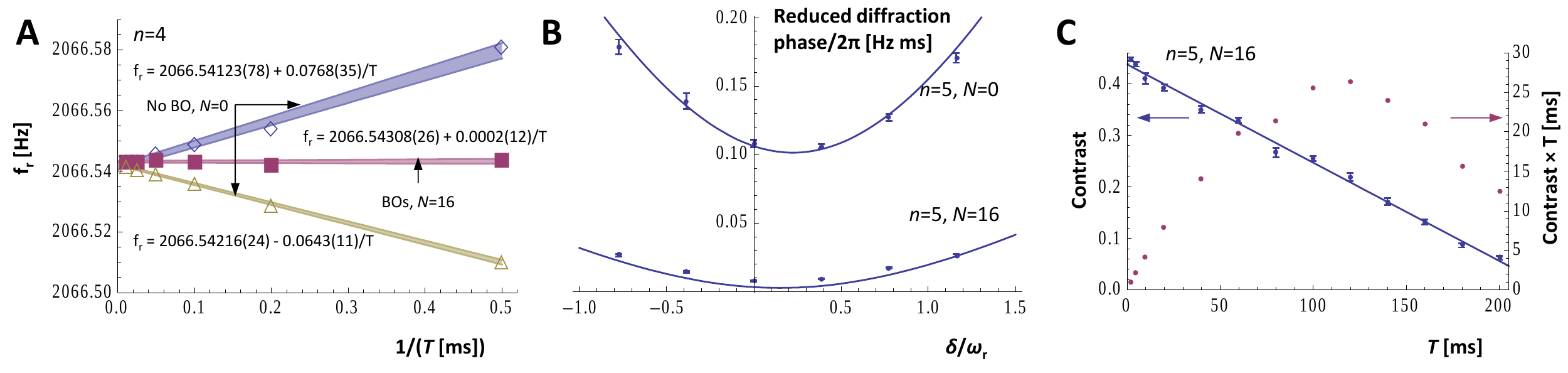,width=\textwidth}
\caption{\label{Data} A: Suppression of the beam splitter phase for $n=4$. The open symbols  show the measured recoil frequency $f_r=\om_r/2\pi$ without Bloch oscillations with Rabi frequencies $\hat \Om_R= 90$ (diamonds) and 100 a.u. (triangles), where 100 a.u. results in a 50\% beam splitter. Least-squares fits determine the diffraction phase as $\Phi_0=0.482(21)$\,mrad and $\Phi_0=-0.403(7)$\,mrad, respectively; shaded areas represent the $1-\sigma$ fit error. Closed symbols are measured at $\hat \Om_R=100$\,a.u. with $N=16$ Bloch oscillations, showing suppression of $\Phi_0$ to 0.0012(8)\,mrad. B: Reduced diffraction phase measured (symbols) and calculated (lines) as function of the two-photon detuning $\delta$. C: Contrast $C(T)$ of the interferometer as function of pulse separation time $T$ with $n=5, N=16$ along with a linear fit. As a measure of the expected sensitivity, we also plot the product $CT$.}
\end{figure*}

Experimentally, we measure the diffraction phase by running the interferometer with different pulse separation times $T$, recording the respective values of $\omega_m$ for which the phase $\Phi$ of the interferometer is zero, and fitting to Eq. (\ref{ommeas}), see Fig. \ref{Data} A. Fig. \ref{Data} B shows the reduced diffraction phase thus measured as function of the two-photon detuning from Bragg resonance for $n=5, N=0$ and $N=16$. To compare experiment with the theory, we used three fit parameters, the peak Rabi frequency of the first pulse pair and the second pulse pair, as well as the offset in the two-photon detuning. The theory curves are averaged over the velocity distribution of the atoms. The intensity per frequency component during the final two Bragg pulses is fitted to be 2.5\% higher than that during the first two pulses. The observed reduced diffraction phase is $\Phi_0= 2\pi \times 0.106(2)$\,mrad. Introducing $N=16$ Bloch oscillations reduces this phase about 10-fold to $2\pi\times 0.0093(5)$\,mrad. 


\begin{figure}
\centering
\epsfig{file=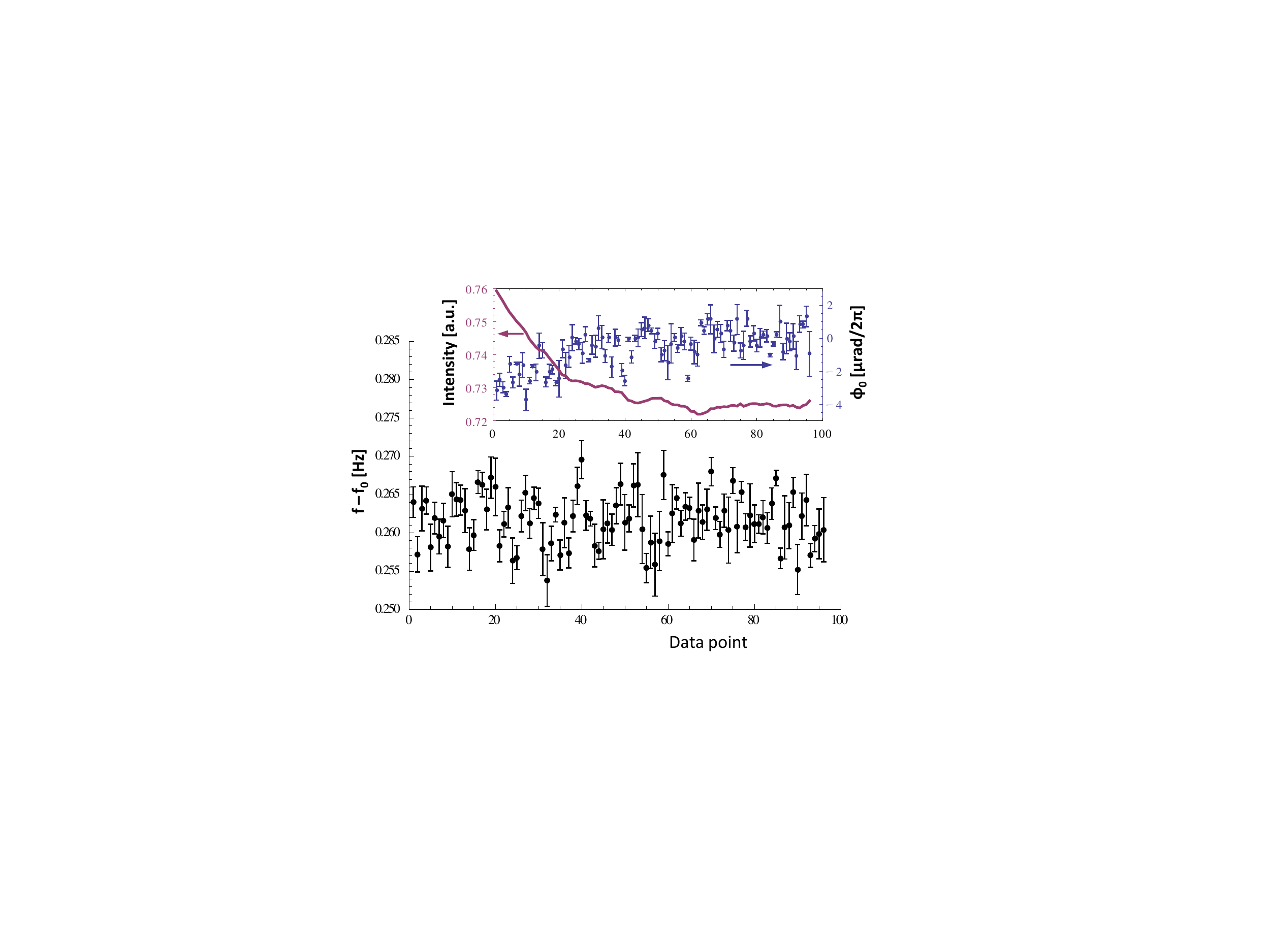,width=0.45\textwidth}
\caption{\label{Longtermdata} 25.3\,hours of data taken with $T=80$\,ms, $n=4$, $N=25$ using feedback to null diffraction phases. Shown is the measured recoil frequency $8(n+N)\om_r$ relative to its average of approximately $2\pi\times 479$\,kHz. The inset shows how the measured diffraction phase is zeroed exponentially upon activating the feedback.}
\end{figure}

An even stronger suppression can be achieved by active feedback to the Bragg pulse intensity to null the diffraction phase. For some diffraction orders, the Bragg pulse duration can be chosen such that zero crossings of the diffraction phase occur for laser pulse intensities close to the ones that give 50\% beam splitting, see Figs. \ref{RBI} B and \ref{SCInew}. Figure \ref{Data} A, solid symbols, shows operation at the zero crossing for 8-photon Bragg diffraction, nulling the reduced diffraction phase. 
We monitor the diffraction phase continuously by alternating between a long and a short pulse separation time $T$ (80 and 5\,ms, respectively) and apply feedback by having the computer adjust the laser pulse energy until this measured diffraction phase is zero, using a proportional-integral servo implemented in software. Figure \ref{Longtermdata} shows the operation of the interferometer with this servo. The reduced diffraction phase is now $2\pi\times 0.030(73)\,\mu$rad, i.e., compatible with zero at a level about 1500 times below what we typically had before this study (as exemplified by Fig. \ref{Data}, B with $N=0$).




Finally, we show that the above methods do not compromise the state-of-the-art sensitivity of the interferometer. Fig. \ref{Data}, C shows the contrast of $n=5, N=16$ interferometers as function of pulse separation time $T$. We reach 45\% contrast--90\% of the theoretical optimum \cite{BraggPRL}--for short $T$. This is significantly better than what we typically achieved before ($\sim 35\%$). The contrast remains nonzero even at $T=200\,$ms and the signal-to noise peaks at $T=130$\,ms. The free-evolution phase differences between the matter waves at these $T$ are respectively $\Phi\simeq 4.4\,$Mrad and $\Phi=2.8$\,Mrad, which are significant improvements relative to previous atom interferometers of any kind that has a nonzero free-evolution phase \cite{Bouchendira2011,CCC}. 
The data in Fig. \ref{Longtermdata} determines the recoil frequency with a resolution of 0.5\,ppb and the fine-structure constant to a state-of-the-art resolution of 0.25\,ppb.

To summarize, we calculated and measured the diffraction phase shift and demonstrated its suppression by introducing Bloch oscillations. Our Ramsey-Bord\'e interferometers combine signal enhancement by multiphoton Bragg diffraction \cite{BraggPRL} and Bloch oscillations \cite{Bouchendira2011}, and suppress vibrations \cite{SCI}, the Coriolis force \cite{Coriolis}, as well as the diffraction phase shift \cite{Buechner,Cronin}. 

We thank Sheng-wey Chiow, Jordan Dudley, Paul Hamilton, Philipp Haslinger, and Erik Urban for discussions and assistance. This work was supported by the David and Lucile Packard Foundation, the National Science Foundation CAREER award number PHY-1056620, and the National Aeronautics and Space Administration grants number NNH13ZTT002N, NNH10ZDA001N-PIDDP, and No. NNH11ZTT001. S.-Y. L. and P.-C. K. acknowledge support from the Singapore National Research Foundation under Grant No. NRFF2013-12

\appendix

\section*{Diffraction phase for various pulse lengths}

In the main text, we use Gaussian pulses $\propto e^{-t^2/2\tau^2}$ with $\tau =0.2\om_r^{-1}$ by default, for both the theory and the experiment. This pulse duration is a good compromise between short pulses (which allow using atoms from a broad velocity distribution) and relatively low-loss Bragg diffraction.

Table \ref{PulseT} shows parameters of a parabolic fit for various pulse durations. In general, the diffraction phase is reduced for long pulses, as expected. This behavior, however, is not uniform. For example, at $n=6$, a tenfold reduction of the dispersion coefficient $\phi_2$ is reduced to $|\phi_2|\sim 25\,$mrad already at $\tau=0.3$. The dispersion also shows several zero crossings, where it can be nulled.

\begin{table}
\centering
\caption{\label{PulseT} Diffraction phase $\tilde \Phi_0=\varphi_0+\varphi_2 \delta^2$ in mrad for various pulse durations $\tau$ at $N=16$. The fits have been made over a range of $-0.4\om_r<\delta <0.4 \om_r$. For larger values of $\delta$, a $\delta^4$ term would have to be included.}
\begin{tabular}{ccccccccccc}\hline\hline
$n$ & $\tau$ & 0.2 & 0.3 & 0.4 & 0.5 & 0.6 & 0.7 & 0.8 & 0.9 & 1 \\ \hline
1 & $\hat \Om_R$ & 0.54 & 0.34 & 0.25 & 0.20 & 0.17 & 0.14 & 0.13 & 0.11 & 0.10  \\
  & $\phi_0$ & -194 & -13.6 & -0.2 & 0.1 & -0.4 & -0.1 & 0.3 & 0.1 & -0.2 \\
  & $\phi_2$ & -29.3 & -10.4 & -1.6 & -0.3 & 0.7 & 0.3 & -0.9 & -0.4 & 0.9 \\ \hline
2 & $\hat \Om_R$ & 1.06 & 0.84 & 0.74 & 0.65 & 0.58 & 0.54 & 0.50 & 0.47 & 0.45 \\
  & $\phi_0$ & -843 & -17.8 & -104 & -126 & -48.5 & -8.2 & -1.1 & -0.4 & 0.0 \\
  & $\phi_2$ & 12.1 & 249 & 352 &  142 & -8.3 & -11.9 & -3.6 & 0.3 & 2.5 \\ \hline
3 & $\hat \Om_R$ & 1.94 & 1.73 & 1.56 & 1.43 & 1.33 & 1.26 & 1.20 & 1.14 & 1.10 \\
  & $\phi_0$ & -224 & -39.8 & -74.1 & -42.7 & -9.1 & 0.9 & 0.5 & 0.8 & 0.4 \\
  & $\phi_2$ & 120 & 318 & 93.0 & 12.7 & 0.8 & 16.0 & 3.2 & -7.0 & 3.1 \\ \hline
4 & $\hat \Om_R$ & 3.25 & 2.98 & 2.74 & 2.56 & 2.42 & 2.32 & 2.23 & 2.16 & 2.09 \\
  & $\varphi_0$ & -13.6 & -51.4 & 65.7 & -1.3 & -2.4 & 1.1 & 2.5 & 3.3 & 3.4 \\
  & $\varphi_2 $ & 238 & 203 & 74.2 & -18.1 & 2.01 & 2.1 & 3.5 & 0.4 & -2.2 \\ \hline
5 & $\hat \Om_R$ & 4.9 & 4.6 & 4.27 & 4.04 & 3.87 & 3.73 & 3.61 & 3.52 & 3.43 \\
  & $\varphi_0$ & -7.8 & -182 & -11.2 & 5.1 & 4.6 & 5.8 & 7.1 & 8.1 & 8.9 \\
  & $\varphi_2$ & 211 & 352 & -18.2 & -4.1 & -10.3 & 10.3 & -24.8 & -9.9 & -10.7 \\ \hline
6 & $\hat \Om_R$ & 7.03 & 6.54 & 6.16 & 6.06 & 5.68 & 5.5 & 5.35 & 5.23 & 5.12 \\
  & $\varphi_0$ & 5.0 & -16.5 & -2.9 & 3.6 & 15.2 & 13.8 & 14.3 & 15.2 & 16.7 \\
  & $\varphi_2$ & 259 & 25.7 & 0.6 & 7.2 & -36.6 & -4.0 & 10.9 & -33.3 & -7.5 \\ \hline
7 & $\hat \Om_R$ & 9.5 & 8.9 & 8.44 & 8.11 & 7.85 & 7.63 & 7.46 & 7.31 & 7.18 \\
  & $\varphi_0$ & -11.1 & -51.2 & 7.5 & 28.3 & 28.6& 25.5 & 28.0 & 31.6 & 32.6 \\
  & $\varphi_2$ & 325 & 146 & -8.8 & -56.7 & -59.4 & -37.6 & -41 & -49.5 & -32.2 \\ \hline
8 & $\hat \Om_R$ & 12.3 & 11.6 & 11.07 & 10.69 & 10.38 & 10.13 & 9.93 & 9.75 & 9.60 \\
  & $\varphi_0$ & 14.4 & 10.0 & 35.2 & 31.0 & 42.7 & 48.5 & 53.8 & 60.4 & 59.0 \\
  & $\varphi_2$ & 143 & -6.4 & -15.9 & 16.1 & -39.1 & -79.6 & -40.0 & -55.5 & -85.7 \\ 
\hline\hline
\end{tabular}
\end{table}

\end{document}